\documentclass[aps,pra,twocolumn,superscriptaddress]{revtex4-1}

\usepackage[english]{babel}
\usepackage{amsmath}
\usepackage{amsthm}
\usepackage{amssymb}
\usepackage{mathrsfs}
\usepackage{booktabs}
\usepackage{color}
\usepackage{hyperref}
\usepackage[ruled,vlined]{algorithm2e}
\usepackage{cases}
\usepackage{dblfloatfix}
\usepackage{multirow}
\hypersetup{
  colorlinks   = true,  
  urlcolor     = blue,  
  linkcolor    = blue,  
  citecolor    = red    
}
\usepackage{mathtools}
\usepackage{natbib}
\usepackage{pgfplots}
\usepackage{subfigure}
\usepackage{url}

\mathtoolsset{showonlyrefs}

\theoremstyle{definition}

\theoremstyle{remark}

\usepackage{scalerel,stackengine}
\stackMath
\newcommand\reallywidehat[1]{%
\savestack{\tmpbox}{\stretchto{%
  \scaleto{%
    \scalerel*[\widthof{\ensuremath{#1}}]{\kern-.6pt\bigwedge\kern-.6pt}%
    {\rule[-\textheight/2]{1ex}{\textheight}}
  }{\textheight}%
}{0.5ex}}%
\stackon[1pt]{#1}{\tmpbox}%
}
\usepackage{comment}
\usepackage{caption}
\usepackage{float}

\SetArgSty{textnormal}

\makeatletter

\newenvironment{widetext2}{%
  \par\ignorespaces
  \setbox\widetext@top\vbox{%
   \vskip15\p@
   \hb@xt@\hsize{%
    \leaders\hrule\hfil
    \vrule\@height6\p@
   }%
   \vskip6\p@
  }%
  \setbox\widetext@bot\hb@xt@\hsize{%
    \vrule\@depth6\p@
    \leaders\hrule\hfil
  }%
  \onecolumngrid
  \let\set@footnotewidth\set@footnotewidth@ii
}{%
  \par
  \twocolumngrid\global\@ignoretrue
  \@endpetrue
}%

\makeatother

\begin{document}

\title{Recapture Probability for anti-trapped Rydberg states in optical tweezers}

\author{R.J.P.T. \surname{de Keijzer}}
\altaffiliation[Corresponding author: ]{r.j.p.t.d.keijzer@tue.nl }

\author{O. \surname{Tse}}

\author{S.J.J.M.F. \surname{Kokkelmans}}

\affiliation{
Department of Applied Physics and Eindhoven Hendrik Casimir Institute, Eindhoven University of Technology, P. O. Box 513, 5600 MB Eindhoven, The Netherlands}

\date{\today}

\begin{abstract}
In a neutral atom quantum computer, the qubits are individual neutral atoms trapped in optical tweezers.
Excitations to Rydberg states form the basis for the entanglement procedure that is at the basis of multi-qubit quantum gates. However, these Rydberg atoms are often anti-trapped, leading to decoherence and atom loss. In this work, we give a quantum mechanical description of the anti-trapping loss rates and determine the recapture probability after Rydberg excitation, distinguishing between having the laser traps turned on and off. We find that there is ample time ($\approx$ 30 $\mu$s, in a Strontium-88 system) needed for the wave functions to expand out off the trap. Therefore, even with traps on, $\approx$ 100\% recapture probabilities can be expected for times in which significant entanglement operations between atoms can be performed. We find that for 2D radial traps with bosonic Strontium-88 atoms, the time in which perfect recapture can be achieved, is of the same order of magnitude for traps on, and off. 
\end{abstract}

\maketitle

\section{Introduction}
\label{sec:introduction}

Neutral atoms trapped in optical tweezers are emerging as a promising platform for scalable quantum computing. Main advantages of these platforms are the demonstrated long coherence time \cite{srlifetime}, the versatility of the atomic arrangements with conservation of entanglement \cite{lukin2}, and generally the contemporary advances in laser cooling techniques for use in atomic clocks, allowing for accurate control of the atoms \cite{review1,review2,review3}. To construct a quantum computer out of such a system, an array of optical tweezers is created with individual alkali or earth-alkaline atoms trapped inside. A consequentially logical choice is then to encode the $|0\rangle$, $|1\rangle$ qubit manifold in the  (meta-)stable states that make up the atomic clock transition. These states are well isolated from the other atoms and have long lifetimes. For bosonic Strontium-88 ($^{88}$Sr), this is the $^1$S$_0\leftrightarrow^3$P$_0$ transition, where the metastable state $^3P_0$ has a lifetime in the order of minutes \cite{srlifetime}. Besides these properties, a quantum computing platform has more stringent requirements on the few-particle level, such as single-qubit control and the ability to entangle neighboring atoms.

\medskip

This entanglement is mediated through auxiliary Rydberg states, which interact via Van der Waals interactions \cite{vdw}. These are electronic structure states of the atom with high principle quantum number $n$ and have significantly lower lifetimes than the clock states due to losses from e.g.\ photo-ionization, spontaneous emission, black body radiation and anti-trapping from the Rydberg states \cite{madhav}. This paper will principally analyze the latter, where anti-trapping is caused by the fact that polarizability of the Rydberg states switches signs compared to the clock states. The resulting light-atom potential becomes concave; actively repelling the atom \cite{expansion}. Therefore, minimizing the time that the atom spends in the Rydberg state is important for optimizing pulse \cite{ownpaper2} or gate robustness \cite{madhav,sven2}.

\medskip

One strategy to avoid such losses is to switch off the trap for the duration of the entanglement procedure, then switch it back on for recapture \cite{madjarov_high-fidelity_2020}. The anti-trapping behavior will then be mitigated because the spatial wave function will evolve under a \textit{Free} potential instead of a concave inverse Gaussian (\textit{IGauss}) potential (resulting from the Gaussian intensity pattern of the optical tweezer \cite{madjarov_2021}). Nevertheless, the atom will still expand under the Free potential and eventually leave the trap after a longer time spent in the Rydberg state \cite{propagator}. Furthermore, there are other challenges when switching off the trap. The blinking on and off of the traps can lead to heating of the atoms in the qubit manifold and thus heat the entire qubit array. On a sequential gate-based platform, this can severely limit the depth of the gate circuit \cite{madjarov_high-fidelity_2020}. One way of avoiding these losses is to use interferometrically generated bottle beam traps to trap both the qubit and Rydberg states \cite{bottletraps}. However, these are technically challenging to create, and generally less deep. 

\medskip

Another possibility is to simply leave the trap on for the Rydberg excitation. It was generally assumed that the anti-trapping would lead to an exponential loss in time of the atom \cite{pagano_error_2022}. However, in recent experiments, it is found that this is not the case \cite{madjarov_high-fidelity_2020, Bluvstein2023, sven2,patent}. This can be understood as, despite the repulsive potential, the atoms need an initialization time to start leaving the trap.

\medskip

In this paper, we shed light on this discussion by researching how recapture probabilities depend on the status of the trap (on or off). The layout of this paper is as follows. Section~\ref{sec:section2} describes the theory behind our quantum mechanical recapture probability model. Here, we first look into the relevant timescales in Sec.~\ref{sec:section2a}. We then consider the potentials and initial states in Sec.~\ref{sec:section2a}. Section~\ref{sec:propagators} describes the evolution methods. In Sec.~\ref{sec:2d} we argue that we can infer 2D results for the trap from our 1D trap calculations. Sec.~\ref{sec:results} shows results for recapture probabilities, based on parameters of a realistic $^{88}$Sr setup. 

\section{Quantum Model of Recapture Probabilities}
\label{sec:section2}

In this section, we discuss the underlying evolution equations and approximations necessary to calculate recapture probabilities. Note that our method is the first fully quantum mechanical model of recapture probabilities considering tweezer status, whereas semiclassical methods have been considered before \cite{classicalmethod, classicalmethods2}.

\begin{figure}[H]
    \centering
    \captionsetup{justification=justified}
    \includegraphics[scale=0.7]{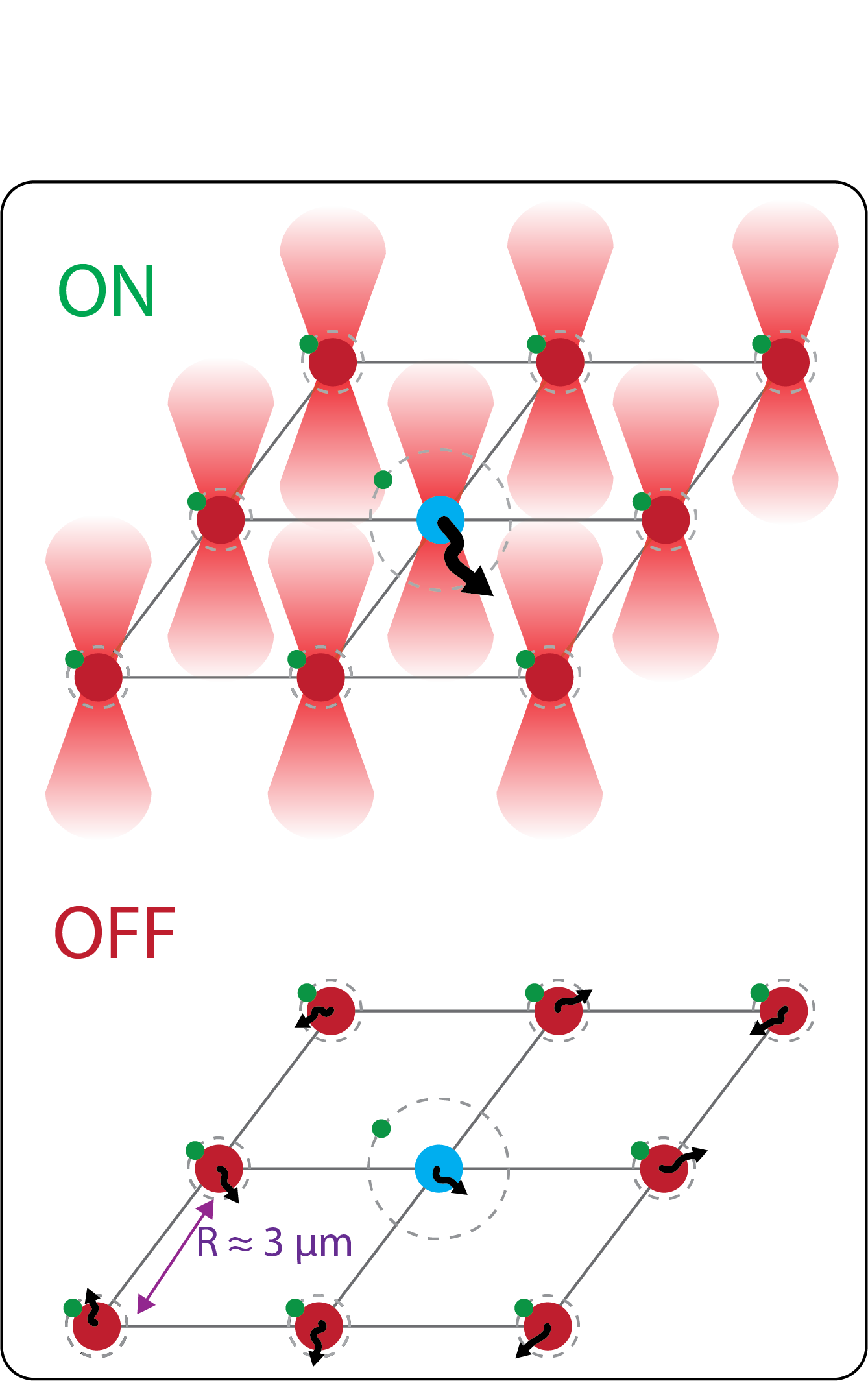}
    \caption{The middle atom in a 3$\times$3 grid of optical tweezer traps, separated from the other atoms by the interatomic separation $R$ of the order of a few micrometers, is excited to the Rydberg state (blue), while all other atoms stay in the qubit manifold (red). When the laser traps are turned on, the excited atom will be actively forced outwards as the tweezer potential becomes repulsive for the Rydberg state. The rest of the atoms stay trapped. When the laser traps are turned off, the drift will be less than when the traps are left on. However, all atoms will expand in this case.}
    \label{fig:my_label}
\end{figure}

\subsection{Relevant Timescales}
\label{sec:timescales}
In this section, we analyze the relevant timescales in the recapture probability problem. When an atom is excited to the Rydberg state, four timescales play a role. First, we consider the excitation time $\tau_{\text{spin}}=1/\Omega_{R}$, with $\Omega_R$ the Rabi frequency. This is the timescale determining how fast one can excite from the qubit manifold to the Rydberg state. In contrast, there is the motional time $\tau_{\text{mot}}=1/\omega$, with $\omega$ the trap frequency, which is the trap motional time determining the speed of evolution of motional states. In current $^{88}$Sr setups, $\Omega_R\approx 10$ MHz and $\omega\approx 20-60$ kHz \cite{madjarov_2021, pagano_error_2022, parameters1, Zhang2020}, showing that the excitation can be considered instantaneous in the timescale of motional development. Therefore, the motional state will not evolve significantly during the excitation from the qubit manifold. Secondly, we consider the interaction time $\tau_{\text{int}}=1/V=-R^6/C_6$, where $C_6$ is the Van der Waals coefficient and $R$ is the interatomic distance defined by the tweezer lattice spacing; and  the time in which we recapture with $\approx 100\%$ certainty $\tau_{\text{recap}}$ (see Eq.~\eqref{eq:survtime}).  When in the Rydberg state, we want to be able to perform sufficient entanglement between the neighboring atoms in the platform, as controlled by $\tau_{\text{int}}$. Thus, we want the recapture time $\tau_{\text{recap}}\gg \tau_{\text{int}}$.

\subsection{Potentials and Initial states}
\label{sec:section2a}
The potential an atom experiences in a Gaussian optical tweezer in cylindrical coordinates $(r,z)$ is given by
\begin{equation}
\label{eq:gausslaser}
    U(r, z)=\frac{U_0}{1+z^2 / z_R^2} \exp \left(\frac{-2 r^2}{w_0^2\left(1+z^2 / z_R^2\right)}\right),
\end{equation}
where $w_0$ is the laser beam waist, $z_R$ the Rayleigh range and the maximal potential $U_0=-\alpha_0 I_0/2c\epsilon_0$ \cite{gauss}. Here, $I_0=2P/\pi w_0^2$ is the central intensity of the laser beam, where $P$ is the total power of the laser. Furthermore, $\alpha_v$ is the polarizability of state $|v\rangle$ \cite{polarizabilitydef} given by
\begin{equation}
\alpha_v(\omega_{\text{laser}})=\frac{2}{3(2 J+1)} \sum_n \frac{\left(E_n-E_v\right)|\langle v|D| n\rangle|^2}{\left(E_n-E_v\right)^2-\omega_{\text{laser}}^2},
\end{equation}
where $J$ is the total angular momentum of $|v\rangle$, $E_i$ is the energy of state $|i\rangle$, $D$ is the electric dipole operator and $\omega_{\text{laser}}$ is the frequency of the laser beam. The polarizability characterizes a state's susceptibility to an electric field. For alkali atoms and highly excited states, polarizabilities can be determined using an atomic structure library \footnote{in this work, the Python library ARC3.0 was used for polarizability calculations \cite{arc}}. For low-energy states of earth-alkaline metals, such as $^{88}$Sr, one needs to resort to more complex methods because of the presence of two or more valence electrons. In this work, we use the methods as described in work by Safronova et al. \cite{polarizabilitysafronova, polarizabilitysafranova2} to calculate polarizability values.

\medskip

\begin{figure}[H]
\captionsetup{justification=justified}
\centering
    \includegraphics[scale=0.85]{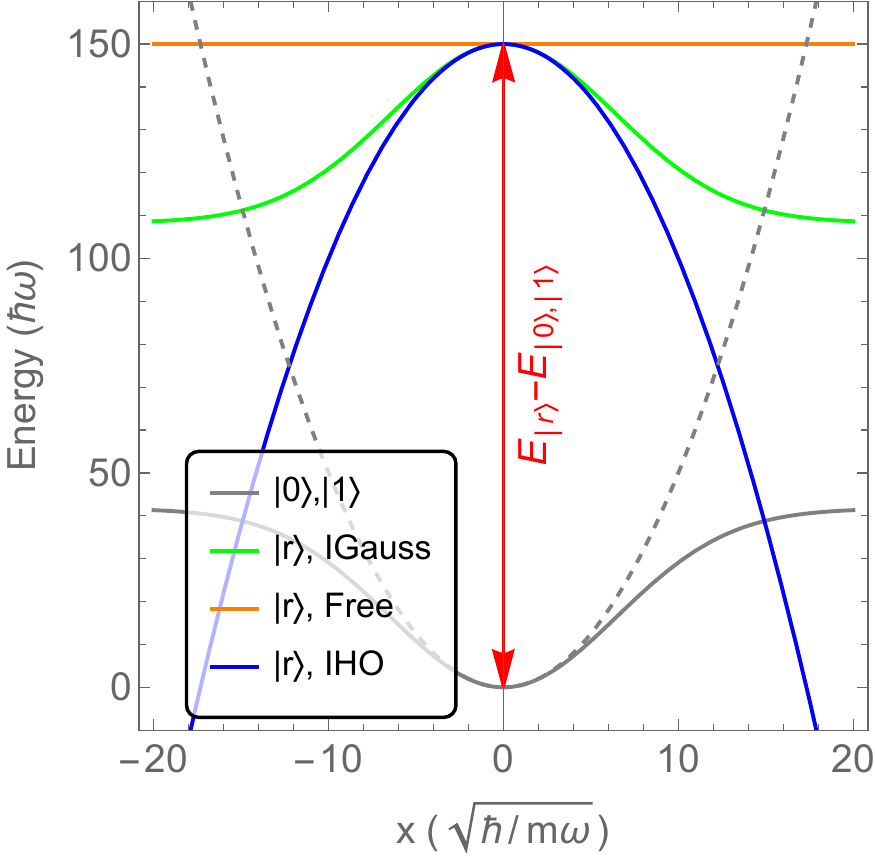}
    \captionsetup{justification=justified}
    \caption{Potentials experienced by the qubit manifold states $|0\rangle$, $|1\rangle$ and the Rydberg state $|r\rangle$, and their (inverse) harmonic oscillator approximations valid for small $|x|$. Difference in state related energies $E_{|r\rangle}-E_{|0\rangle,|1\rangle}$ is not to scale.}
    \label{fig:potentialsinverted}
\end{figure}

We consider a $^{88}$Sr based quantum computer using the states forming the clock transition, $5s^2\, ^1S_0$ and $5s5p\, ^3P_0$ as our qubit states. For these states, a magic wavelength has been identified at 813.4 nm, where the polarizabilities of both states are equal to 287 a.u. \cite{polarizabilitysafronova}. Operating the traps at a magic wavelength is preferable, as it eliminates phase build-up between the qubit states due to AC Stark shifts. For the Rydberg state, we find a polarizability of the same magnitude but different sign \cite{patent}. Thus, the Rydberg state experiences an inverted potential compared to the qubit states, when the laser is turned on, as illustrated in Fig.~\ref{fig:potentialsinverted}. Even when the laser is turned off, and the atom subsequently experiences no potential, the wave function will expand under the Free particle propagator. This means that when an atom is excited to the Rydberg state, it will eventually leave the trap and can no longer be recaptured when it is deexcited back to the qubit manifold. In the rest of this section we describe our model for calculating recapture probabilities for when we excite an atom from the qubit manifold to the Rydberg state, let the wave function expand for a set time $t$, and deexcite it back to the qubit manifold.

\medskip
Since recapture is only sensitive to radial motion \cite{thompsonsensitive}, only the radial directions of the trap are considered, we thus take $z=0$. Furthermore, in Sec.~\ref{sec:2d}, we will argue that we can infer the 2D (radial) recapture probabilities from 1D results. In the rest of this section, we thus consider a 1D optical tweezer trap. Based on Eq.~\eqref{eq:gausslaser}, we consider the potentials
\begin{equation}
\begin{aligned}
\label{eq:potentials}
V_{\text{HO}}(x)&=\frac{1}{2}m \omega^2x^2, \quad V_{\text{Free}}(x)=0,\\
V_{\text{Gauss}}(x)&=-U_0+ U_0 \exp(-2x^2/w_0^2),
\end{aligned}
\end{equation}
for respective HO, Gauss, and the Free potential. We also define $V_{\text{IHO}}=-V_{\text{HO}}$ and $V_{\text{IGauss}}=-V_{\text{Gauss}}$ for the inverse HO and inverse Gauss potentials. The harmonic oscillators in this case are approximations to the Gaussian potentials for small $|x|$ and will often be used as such, see Fig.~\ref{fig:potentialsinverted}. Using the approximation $\exp(y)\approx 1+y$ in Eq.~\eqref{eq:potentials}, the values of $\omega$ are related to $U_0, z_R$ and $w_0$ as
\begin{equation}
\label{eq:frequencies}
\omega_{x, y} =\sqrt{\frac{4 |U_0|}{m w_0^2}},\qquad
\omega_z =\sqrt{\frac{2 |U_0|}{m z_R^2}}.
\end{equation}

When the atom is in the Rydberg state, its wave function $|\psi(x,t)\rangle$ will evolve under the time-dependent 1D Schrödinger equation as
\begin{equation}
\begin{aligned}
    i\hbar \partial_t|\psi(x,t)\rangle=-\frac{\hbar^2}{2m}\hbar^2\partial_{xx}&|\psi(x,t)\rangle+V(x)|\psi(x,t)\rangle,\\
    |\psi(x,0)\rangle&=|\psi_0\rangle.
    \end{aligned}
\end{equation}
Based on Eq.~\eqref{eq:potentials}, we transform our coordinates as
\begin{equation}
x\rightarrow \tilde{x}\cdot x_0,\, t\rightarrow \tilde{t} \cdot t_0,\quad  \text{with} \, x_0=\sqrt{\hbar/m\omega},  t_0=1/\omega, 
\end{equation}
to give the dimensionless equation
\begin{equation}
\begin{aligned}
    i\partial_{\tilde{t}}|\psi(\tilde{x},\tilde{t})\rangle=-\frac{1}{2}\partial_{\tilde{x}\tilde{x}}|\psi(\tilde{x},&\tilde{t})\rangle+\frac{1}{\hbar \omega}V(\tilde{x})|\psi(\tilde{x},\tilde{t})\rangle,\\ |\psi(\tilde{x},0)\rangle&=|\psi_0\rangle.
\end{aligned}
\end{equation}

For convenience, we will drop the tildes on the transformed coordinates. For Gaussian laser traps, we also define the edge of the trap $X_{\text{edge}}$ as
\begin{equation}
\label{eq:trapedge}
\begin{aligned}
  & \frac{ V_{\text{Gauss}}(X_{\text{edge}})}{V_{\text{Gauss}}(0)}\approx 0.01\\ \Rightarrow & X_{\text{edge}}\approx 3.035 \sqrt{\frac{|U_0|}{\hbar\omega}}.
\end{aligned}
\end{equation}

In contemporary $^{88}$Sr tweezer setups, trap depths $U_0$ of several hundreds $\mu$K can be achieved. However, when exciting to the Rydberg state, the traps are often adiabatically lowered to minimize trap scattering (which is especially important in $^{88}$Sr) and to increase the recapture probability \cite{madjarov_2021}.

\medskip 

As the atoms in the trap are cooled close to the ground state, the initial state of the system can be approximated by the density matrix of a quantum harmonic oscillator at temperature $T$ \cite{kauffman}, given by
\begin{equation}
\label{eq:densitymatrix}
    \rho_{T}(0)=\frac{1}{Z}\sum_{n=0}^\infty e^{-\beta E_{\text{HO},n}}|\psi_{\text{HO},n}\rangle\langle\psi_{\text{HO},n}|,
\end{equation}
where $\beta=1/\text{k$_\text{B}$}T$, $E_{\text{HO},n}=\hbar \omega (n+\frac{1}{2})$, $Z=\sum {e^{-\beta E_{\text{HO},n}}}$ and 
\begin{equation}
    |\psi_{\text{HO},n}\rangle=\frac{1}{\sqrt{2^n n !}}\left(\frac{m \omega}{\pi \hbar}\right)^{1 / 4} e^{-\frac{m \omega x^2}{2 \hbar}} H_n\left(\sqrt{\frac{m \omega}{\hbar}} x\right),
\end{equation}
is the $n$th eigenstate of the HO potential, with $H_n$ the $n$th Hermite polynomial. Let $\rho_T(t)$ evolve under an arbitrary Schr\"{o}dinger equation with initial state $\rho_{T,0}$, and let $|\psi_{\text{HO},n}(t)\rangle$ denote the solution to this same equation with initial state $|\psi_{\text{HO},n}\rangle$. Then the expectation value of an observable $\hat{O}$ w.r.t.\ $\rho_T(t)$ can, by orthogonality of the $|\psi_{\text{HO},n}\rangle$ be calculated as
\begin{equation}
    \text{Tr}[\rho_T(t)\hat{O}]=\frac{1}{Z} \sum_{n=0}^\infty e^{-\beta E_{\text{HO},n}}\langle\psi_{\text{HO},n}(t)|\hat{O}|\psi_{\text{HO},n}(t)\rangle.
\label{eq:thermal}
\end{equation}

\begin{figure}[H]
    \centering
    \includegraphics[scale=0.85]{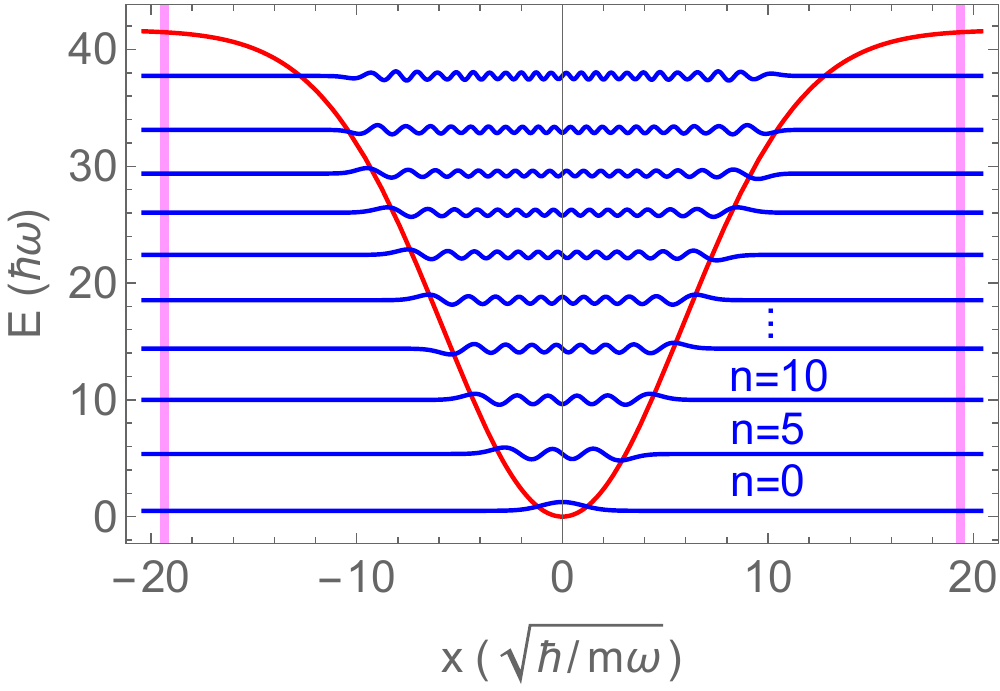}
    \caption{Potential and approximated eigenvalues for non-inverted 1D Gauss potential, with $U_0=50$ $\mu$K and $\omega=25$ kHz. The edges of the trap are colored in purple. Plotted per 5 eigenfunctions.}
    \label{fig:boundstates}
\end{figure}

We want to excite the atom to the Rydberg state, where it will experience a potential $V_{\text{IGauss}}$ (or 0 if we switch off the trap) for a time $t$. After this, we de-excite the atom back to the qubit manifold and determine the Franck-Condon overlap with the bound states of the potential $V_{\text{Gauss}}$ (see Eq.~\eqref{eq:overlap}). To do so, we have to determine the bound states $|\psi_{\text{Gauss},n}\rangle$ (of which there are only finitely many, in contrast to the HO potential \cite{finitebound}). We would have to solve the equation  
\begin{equation}
   -\frac{1}{2}\partial_{xx}|\psi(x,t)\rangle+\frac{U_0}{\hbar \omega}\Bigl(1-e^{-2x^2x_0^2/w_0^2}\Bigr)|\psi(x,t)\rangle=E,
\end{equation}
for energies $E<0$. However, no known analytic solution is known for the Gauss potential \cite{nosolution}. As mentioned before, for $|x|\ll1$, $V_{\text{Gauss}}$ resembles $V_{\text{HO}}$. Therefore, we approximate solutions to the normal Gauss potential as linear combinations of harmonic oscillator solutions (see Fig.~\ref{fig:boundstates}), as
\begin{equation}
    |\psi_{\text{Gauss},n}\rangle=\sum_{i=0}^K\alpha_i|\psi_{\text{HO},n}\rangle,
    \label{eq:boundstatesHO}
\end{equation}
for some $K\in N$. As the Gauss potential resembles the HO potential for $|x|\ll 1$, and the energies in the HO potential increase as $E_n=\hbar \omega(n+1/2)$, $U_0/\hbar\omega$ is a rough estimate of the number of bound states, on which the choice of $K$ is based.\\

The eigensystem of the non-inverted Gauss Hamiltonian in the basis of the first $K$ eigenstates of the HO is calculated by diagonalizing the matrix $\mathcal{A}$ with matrix elements [$\mathcal{A}$]$_{m,n}=\langle \psi_{\text{HO},m}|-1/2\partial_{xx}+V_{\text{Gauss}}(x)|\psi_{\text{HO},n}\rangle$. The bound state approximations are then given by those states with eigenvalue smaller than zero.

\begin{figure}[H]
    \centering
    \includegraphics[scale=0.75]{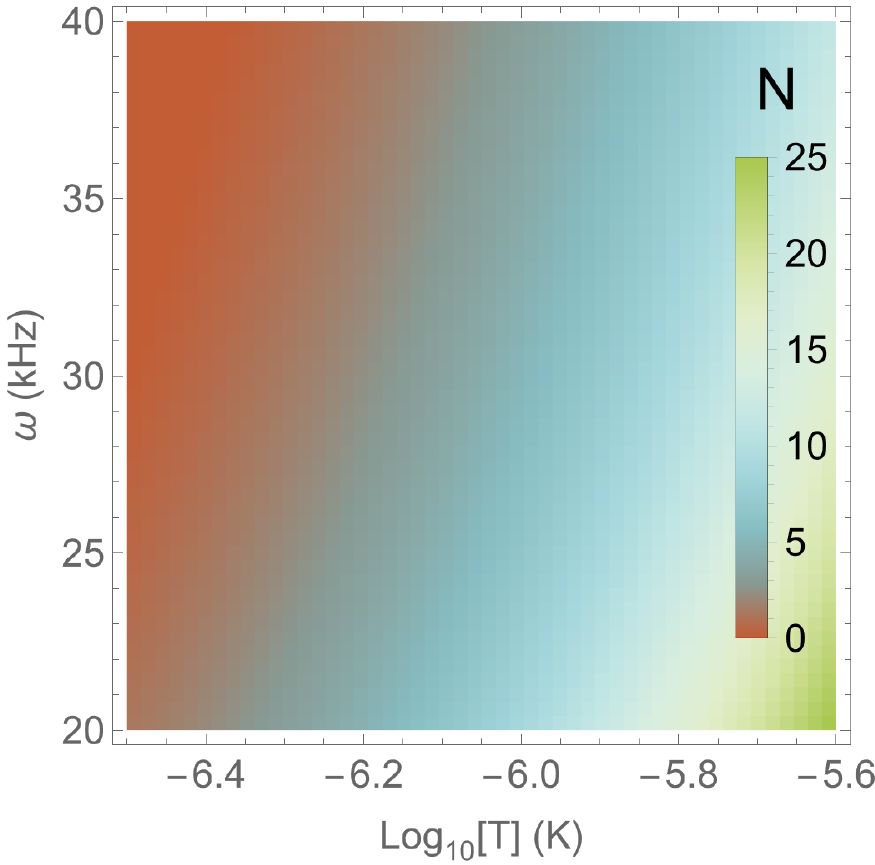}
    \caption{Levels $N$ at which $Z_N/Z>0.99$ for varying temperatures $T$ and frequencies $\omega$. For low $T$ and high $\omega$, it is enough to only take the first few HO eigenstates into account for the initial state.}
    \label{fig:tempvsomega}
\end{figure}

At low temperatures, the thermalized state of Eq.~\eqref{eq:densitymatrix} will only have significant contributions of the lowest energy eigenstates. Therefore, we have to look at the values of 
\begin{equation}
\begin{aligned}
    \frac{Z_N}{Z}&=\frac{\sum_{n=0}^N \exp(-\beta E_n)}{Z}\\
    &=2\exp\left(-\frac{\hbar\omega (1 + N)}{2 k_b T}\right)\sinh\left(\frac{\hbar\omega (1 + N)}{2 k_b T}\right),
\end{aligned}
\end{equation}
for a decent cut-off of the series expansion. Here, $Z_N$ is the cut-off partition function, quantifying how much of the thermalized state is in the first $N$ levels. For a specific temperature $T$, we choose our $N$ as the minimal value such that $Z_N/Z>0.99$. Figure~\ref{fig:tempvsomega} shows these values of $N$ as a function of $T$ and $\omega$.

\subsection{Propagators and recapture probabilities} 
\label{sec:propagators}
With the initial state and bound states defined, the 
next step is to calculate the evolution of the state under various potentials. The highly dispersive behavior of wave functions under concave potentials \cite{griffiths} makes it difficult to numerically solve the time-dependent Schr\"{o}dinger equation.\\

Interestingly enough, there exist analytic propagators for the Free and IHO solutions \cite{propagatorsIHO}, given by
\begin{equation}
\begin{aligned}
    &K_{\text{Free}}\left(x; x^{\prime} ; t- t^{\prime}\right)\\
    &=\frac{1}{2 \pi} \int_{-\infty}^{+\infty} \exp\big(i k\left(x-x^{\prime}\right)\big) \exp\left(-\frac{i \hbar k^2 (t-t^{\prime})}{2 m}\right) d k \\
    &=\left(\frac{m}{2 \pi i \hbar (t-t^{\prime})}\right)^{\frac{1}{2}} \exp\left(-\frac{m\left(x-x^{\prime}\right)^2}{2 i \hbar (t-t^{\prime})}\right),
\end{aligned}
\end{equation}
for the Free Hamiltonian, and
\begin{equation}
\begin{aligned}
&K_{\text{IHO}}\left(x;x^{\prime}; t- t^{\prime}\right)\\
&=\sqrt{\frac{\omega}{2 \pi i \hbar \sinh \big(\omega (t-t^{\prime})\big)}} \exp \left(\frac{i}{\hbar} S(x,t,x^{\prime},t^{\prime})\right),\\
&S(x,t,x^{\prime},t^{\prime})=\frac{\omega\left[\cosh (\omega (t-t^{\prime}))\left(x^2+x^{\prime 2}\right)-2 x x^{\prime}\right]}{2 \sinh \big(\omega (t-t^{\prime})\big)}
\end{aligned}
\end{equation}
for the Hamiltonian with the inverse HO potential.

\medskip

For $|\psi(x,0)\rangle=|\psi_{\text{HO},0}(x)\rangle$, this gives
\begin{equation}
\begin{aligned}
|\psi_{\text{Free}}(x,t)\rangle&=-\frac{(-1)^{3/4}}{\pi^{1/4} \sqrt{t-i}} \exp\left(\frac{i x^2}{2 (t-i)}\right),\\
    |\psi_{\text{IHO}}(x, t)\rangle&=\pi^{-1 / 4} \Gamma(t)^{-1 / 2} \exp \left(\frac{i S\left(x, t, 0, 0\right)}{\hbar}\right),\\
    \times &\exp \left(-\frac{i \omega}{2 \hbar \sinh (\omega t)}\frac{x^2}{\Gamma(t)}\right),\\
\text{with}\quad \Gamma(t)&=\cosh (\omega t)+i \frac{\hbar}{\omega} \sinh (\omega t).
\end{aligned}
\end{equation}
Expressions for other initial states can be constructed similarly. In the literature, one often defines the survival probability $P_{\text{init}}(t)$ \cite{recapprob1, recapprob2} as
\begin{equation}
    P_{\text{init}}(t):=|\langle \psi(t)|\psi_0\rangle|^2,
\end{equation}
i.e. the overlap with the initial state at time $t$. One can show that at $t=0$, its first derivative is equal to zero (see Appendix~\ref{sec:decayrate}). Looking at the second derivative at $t=0$ then gives an indication of the decay rate under a certain potential. We call this term $\ddot{P}_{\text{init},V}(0)$ the \textit{initial quantum spread} under potential $V$. The expression for this is derived in Appendix~\ref{sec:decayrate} and is given by 
\begin{equation}
     \ddot{P}_{\text{init},V}(0)=-2\text{Cov}_{\psi_0}\left(V-\frac{1}{2}\frac{\partial^2}{\partial x^2},V-\frac{1}{2}x^2\right).
     \label{eq:decayrates}
\end{equation}

We can further extend this idea to looking at the overlap with all bound states, called the {\em recapture probability}:
\begin{equation}
\label{eq:overlap}
    P_{\text{recap}}(t):=\sum_{n=0}^N|\langle \psi(t)|\psi_{\text{Gauss},n}\rangle|^2,
\end{equation}
which can be interpreted as the probability of recapturing the atom in the trap after deexciting it back to the qubit manifold after it has evolved for a time $t$. Using the analytic expressions for $|\psi(t)\rangle$ and the HO approximation of the bound states as in Eq.~\eqref{eq:boundstatesHO}, we can calculate $P_{\text{recap}}(t)$ analytically with the integral expression given in Appendix~\ref{sec:integrals}. Based on this Eq.~\eqref{eq:overlap}, we define 
\begin{equation}
\label{eq:survtime}
    \tau_{\text{recap}}:=\max_{t>0}P_{\text{recap}}(t)>1-10^{-4},
\end{equation}
where the $4$ in the exponent is taken to match recently achieved fidelity figures for single qubit gates in Rydberg architectures \cite{fidelity1,fidelity2,review2}. 

\subsection{2D considerations}
\label{sec:2d}
Having defined the model for a 1D trap, the radial 2D case needs to be considered. When looking at the 2D case, the following approximation can be made. If the trap is deep enough, the $x$ and $y$ directions will be independent, and the bound states will factor as $|\psi_{\text{Gauss},n_1,n_2}(x,y)\rangle=|\psi_{\text{Gauss},n_1}(x)\rangle|\psi_{\text{Gauss},n_2}(y)\rangle$. If the expanding wave function then also were to factor as $|\psi_{\text{2D}}(x,y,t)\rangle=|\psi_{\text{1D}}(x,t)\rangle|\psi_{\text{1D}}(y,t)\rangle$ then the recapture probability would become
\begin{equation}
\begin{aligned}
    &P_{\text{recap,2D}}(t)=\sum_{n_1,n_2}|\langle \psi_{\text{2D}}(t)|\psi_{\text{Gauss},n_1,n_2}\rangle|^2\\
    &=\sum_{n_1}|\langle \psi_{\text{1D}}(t)|\psi_{\text{Gauss},n_1}\rangle|^2\sum_{n_2}|\langle \psi_{\text{1D}}(t)|\psi_{\text{Gauss},n_2}\rangle|^2\\
    &=P_{\text{recap,1D}}(t)^2.
    \end{aligned}
\end{equation}
Clearly, the bound functions of the 2D case do not exactly factor as such, as a combination of a barely unbound state in $x$ with a deep bound state in $y$ can still have negative energy and thus be bound. Furthermore, the $x$ and $y$ axis are not entirely independent since a state's projections on the x- and y-axis might be individually bound, but the total state can be unbound, because it extends significantly in a diagonal direction. However, as the wave function extends quickly when reaching the end of the trap, we hypothesize that this approximation is valid for deep traps. To verify this, we will also solve the 2D Schrödinger equation in radial coordinates as 
\begin{equation}
\begin{aligned}
i\partial_t |\psi(r,\phi,t)\rangle 
&=-\frac{1}{2}\Delta_{r,\phi}|\psi(r,\phi,t)\rangle+V(r)|\psi(r,\phi,t)\rangle\\
\text{with}\quad \Delta_{r,\phi} &= \left(\frac{1}{r}\frac{\partial}{\partial r}\left(r\frac{\partial}{\partial r}\right)+\frac{1}{r^2}\frac{\partial^2}{\partial \phi^2}\right).
\end{aligned}
\end{equation}

Bound states for the 2D Gaussian are constructed, similarly as in the 1D case, as linear combinations of radial 2D HO solutions given by
\begin{equation}
   |\psi_{\text{HO},n,l}(r,\phi)\rangle= \frac{1}{\sqrt{2 \pi}}  
  \sqrt{\frac{2 \alpha!}{(\alpha +|l|)!}}e^{-\frac{r^2}{2}}r^{|l|}
   L_\alpha^{(|l|)}(r^2)e^{il\phi},
   \label{eq:radialboundstates}
\end{equation}
where $l\in\{-2n, -2n+2, ... , 2n\}$, $\alpha=(n-|l|)/2$ and $L_{\alpha}^{(|l|)}$ is the generalized Laguerre polynomial. We will always assume a radial initial wave function such that the evolved state $|\psi(t)\rangle$ will not have an angular dependence at all times $t$. Therefore, we only need to calculate the overlap with bound states that also do not have an angular dependence. If the initial wave function does have a radial dependence (for instance because of non-zero temperatures) the results won't change significantly as thermalized states are considered (see Eq.~\eqref{eq:thermal}) and the overlaps and energy differences for angular excited states do not differ substantially from the radially symmetric states (see Eq.~\eqref{eq:laguerreintegral}). 

\section{Results}
\label{sec:results}

This section will show results for the recapture probabilities of atoms for tweezers on and off and varying tweezer parameters. In all calculations, we will use the parameter values as in Tab.~\ref{tab:givenvalues} \cite{madjarov_2021,madjarov_high-fidelity_2020,parameters1,pagano_error_2022,Zhang2020}, unless specified otherwise. From these parameters, we get $\tau_{\text{int}}=4.7$ ns. We desire to have recapture time $\tau_{\text{recap}}\gg\tau_{\text{int}}$.

\begin{table}[t]
\begin{tabular}{lcccc}
\hline \hline Trap frequency & $\omega$  & 25 & kHz & \cite{madjarov_2021, madjarov_high-fidelity_2020} \\
Trap depth & $U_0$ & 50 & $\mu$K & \cite{madjarov_2021, parameters1}\\
Mass & m & 87.90 &  a.m.u. \\
Temperature & $T$ & 730 & nK & \cite{madjarov_2021, parameters1} \\
VdW coefficient & $C_6/h$ & -154 & GHz/$\mu$m$^6$ & \cite{pagano_error_2022}\\
Interatomic distance & $R$ & 3 & $\mu$m 
& \cite{madjarov_2021, Zhang2020}\\
\hline \hline
\end{tabular}
\caption{Parameters used in simulation and calculations.}
\label{tab:givenvalues}
\end{table}

We construct the bound state for the realistic parameters $U_0=50$ $\mu$K and $\omega=25$ kHz, which gives $U_0/\hbar \omega\approx 41$, so we take $K=55$ and find $48$ bound states (see Fig.~\ref{fig:boundstates}) according to the linear approximation method described in Sec.~\ref{sec:section2a}. For $n<10$ we then have that $\alpha_n\approx1$ and thus $|\psi_\text{Gauss},n\rangle\approx |\psi_{\text{HO},n}\rangle$, showing that the HO eigenfunctions are good approximations of the non-inverted Gauss eigenfunctions for these low-lying states.

\begin{widetext2}
    \begin{minipage}[b]{\linewidth}
\begin{figure}[H]
\begin{centering}
\captionsetup{justification=raggedright}
    \begin{minipage}[t]{.444\textwidth}
         \includegraphics[width=\columnwidth]{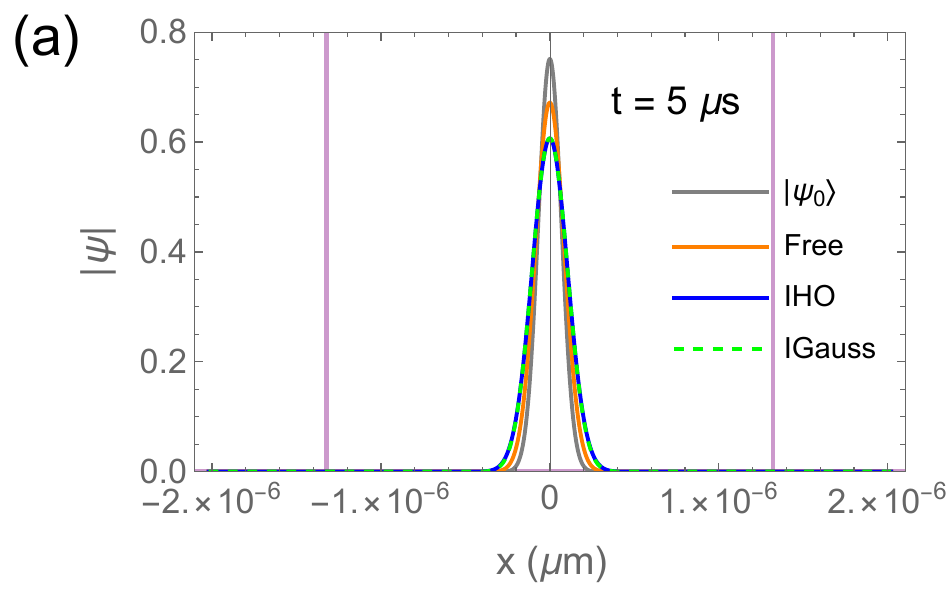}
    \end{minipage}%
    \begin{minipage}[t]{.444\textwidth}
         \includegraphics[width=\columnwidth]{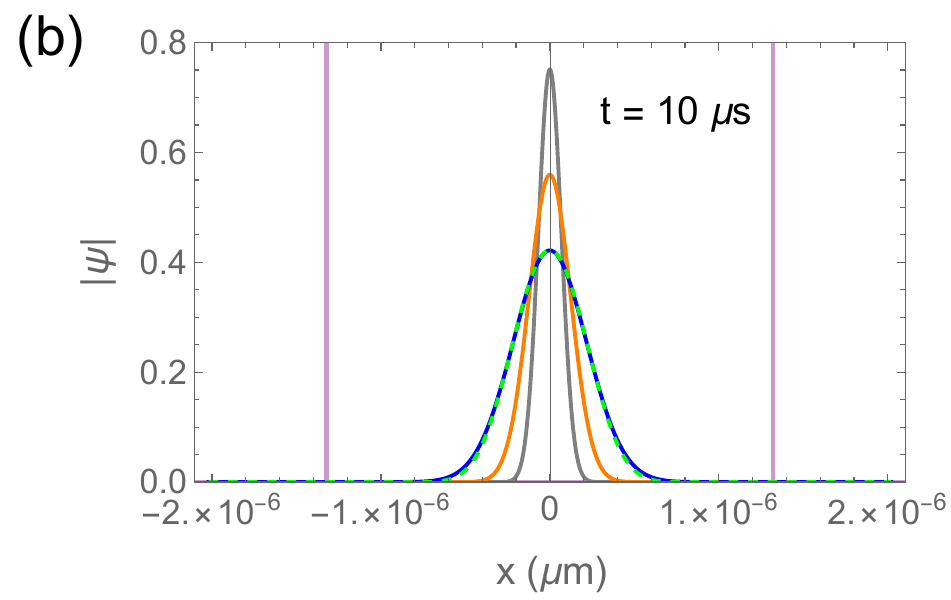}
    \end{minipage}%
    \\
    \begin{minipage}[t]{.444\textwidth}
         \includegraphics[width=\columnwidth]{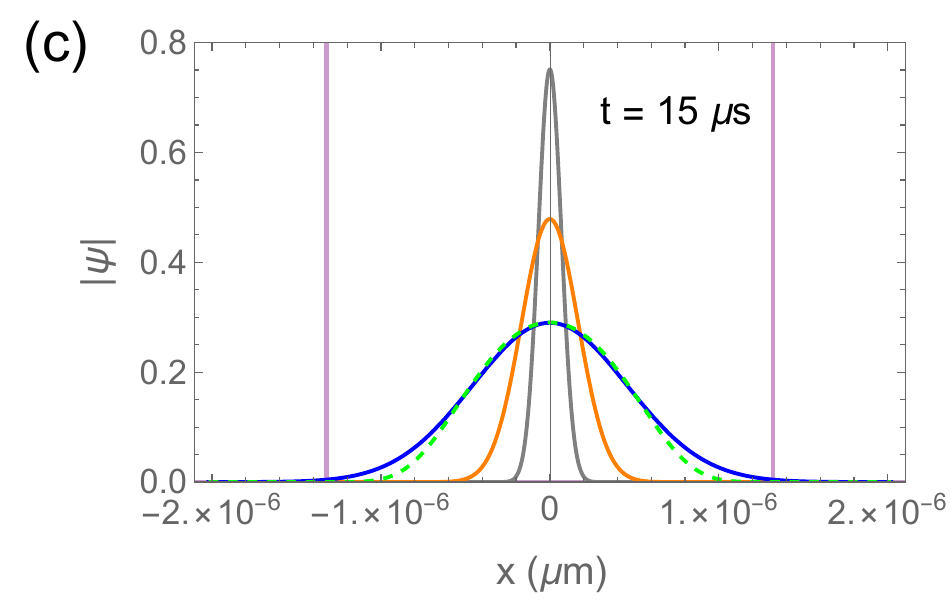}
    \end{minipage}%
    \begin{minipage}[t]{.444\textwidth}
         \includegraphics[width=\columnwidth]{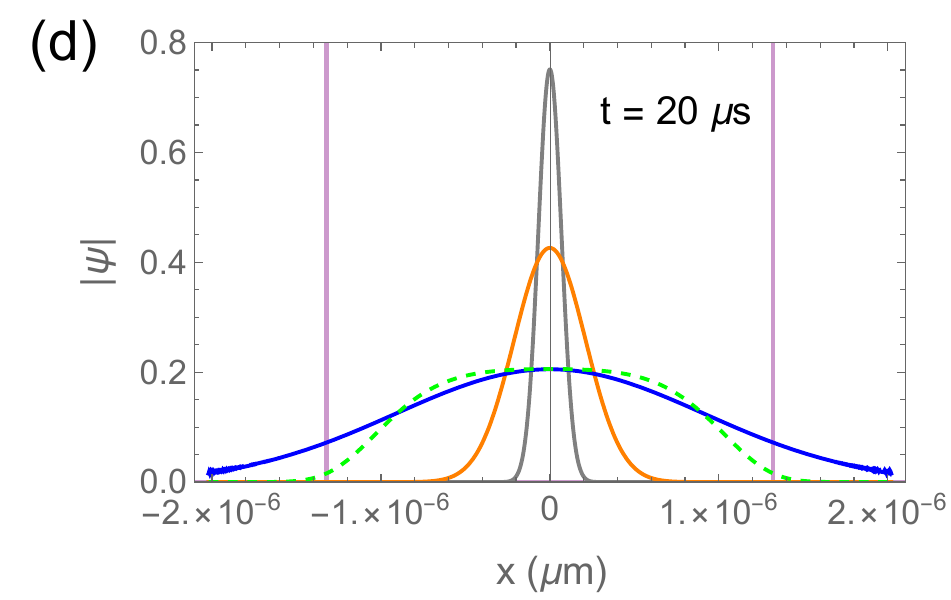}
    \end{minipage}%
\caption{Evolution of the $|\psi(x,0)\rangle=|\psi_{\text{HO},0}\rangle$ initial state for different 1D potentials, with $U_0=50$ $\mu$K and $\omega=$25 kHz. For times $t<15$ $\mu$s the IHO approximation agrees with the IGauss solution. Once the atoms start leaving the trap, numerical errors start to occur (see edges of (d)) and the IHO approximation is no longer valid. The edges of the trap are indicated in purple.}
    \label{fig:stateevolution}
    \end{centering}
\end{figure}
    \end{minipage}
\end{widetext2}

\textcolor{white}{.}
\newpage
\textcolor{white}{.}
\newpage

Figure~\ref{fig:stateevolution} shows the (numerically calculated) evolution of a 1D $|\psi_{\text{HO},0}\rangle$ initial state under various potentials, with $U_0=50$ $\mu$K and $\omega=25$ kHz. We see that the IHO potential is a good approximation of the IGauss evolution for times $t<15$ $\mu$s. Note that numerical errors start to occur at times $t>20$ $\mu$s. This is due to the high dispersivity caused by the concave potentials \cite{griffiths}, which makes a Dirichlet or Von Neumann boundary problem \cite{boundary} hard to solve, highlighting the importance of the analytic propagator expressions in Sec.~\ref{sec:propagators}. 

\medskip

Figure~\ref{fig:survivalrates} shows the recapture probabilities of the $U_0=50$ $\mu$K and $\omega=25$ kHz trap for the Free, IHO and IGauss potentials. Note that because of numerical error, the IGauss evolution is only fully accurate up to 30 $\mu$s. From this, we again confirm that the IHO potential is a good approximation to the IGauss evolution for low enough evolution times. From Eq.~\eqref{eq:decayrates}, we get initial quantum spreads of $\ddot{P}_{\text{init,IHO}}(0)=-1$ and $\ddot{P}_{\text{init,IGauss}}(0)=-0.98$. The inset of Fig.~\ref{fig:survivalrates}, indeed confirms these results by showing that the initial loss rates for an IHO potential are higher than for a IGauss potential. Furthermore, as hypothesized in Sec.~\ref{sec:introduction}, a plateau of $\approx100\%$ recapture probabilities exists for repelling potentials. We see values of $\tau_{\text{recap,IGauss}}/\tau_{\text{recap,Free}}\approx 0.33$. When switching off the traps, all atoms will expand under the Free potential, whereas, when the traps are kept on, only one atom will expand under the IGauss potential. These results would then indicate that the overall heating would be significantly lower when the traps are kept on. The blinking heating and control issues, persisting when repeatedly switching the traps on and off, further support the strategy of leaving the traps on \cite{madjarov_2021}.

\medskip

To further investigate this statement, we vary the trap parameters $U_0$ and $\omega$, and calculate $\tau_{\text{recap,IGauss}}$ and $\tau_{\text{recap,Free}}$. The results in Fig.~\ref{fig:trecapgauss} show that $\tau_{\text{recap}}$, increases with decreasing $\omega$ and with increasing $U_0$. For $\omega$, this behavior is logical as traps become tighter for increasing $\omega$ (see Eq.~\eqref{eq:frequencies}), resulting in fewer bound states. For $U_0$, the traps become deeper, resulting in faster expansion, but also more bound states, which for these range of parameters seems to be the dominating factor. We see that the contour lines in Fig.~\hyperlink{fig:trecapgaussa}{\ref{fig:trecapgauss}a} are approximately straight lines for our considered range of parameters. We can therefore take the trap edge $X_{\text{edge}}$ of Eq.~\eqref{eq:trapedge}, to be a predictor of the recapture time $\tau_{\text{recap}}$. Figure~\hyperlink{fig:trecapgaussb}{\ref{fig:trecapgauss}b} shows the values of the ratio $\tau_{\text{recap,IGauss}}/\tau_{\text{recap,Free}}$ for the range of parameters, and we see that the values lie in the interval $[0.3,0.6]$, indicating that the recapture times stay in the same order of magnitude.

\begin{widetext2}
    \begin{minipage}[b]{\linewidth}
\begin{figure}[H]
\begin{centering}
\captionsetup{justification=justified}
    \begin{minipage}[t]{.47\textwidth}
         \includegraphics[width=\columnwidth]{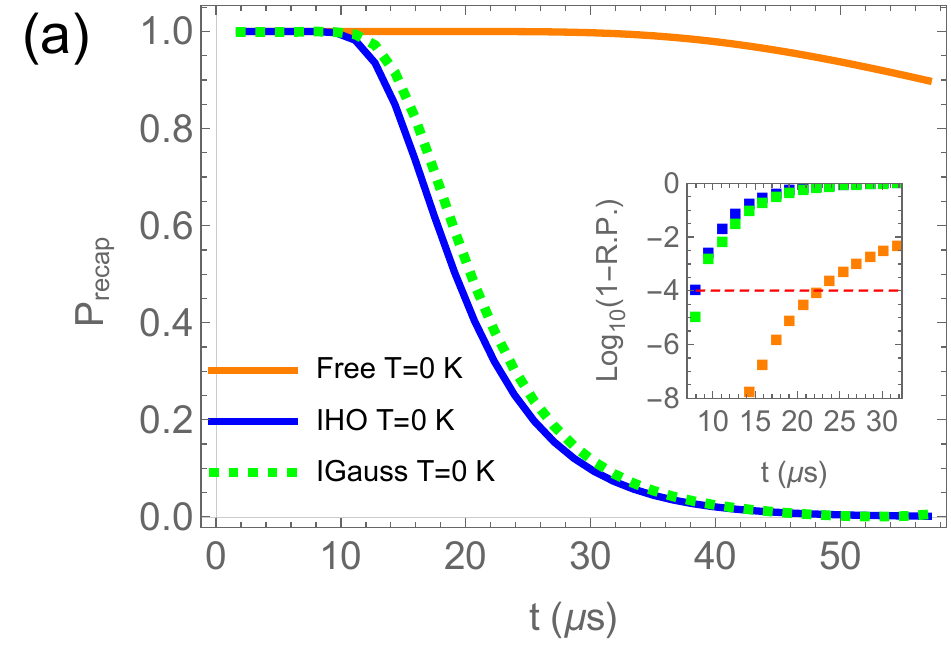}
    \end{minipage}%
    \begin{minipage}[t]{.47\textwidth}
         \includegraphics[width=\columnwidth]{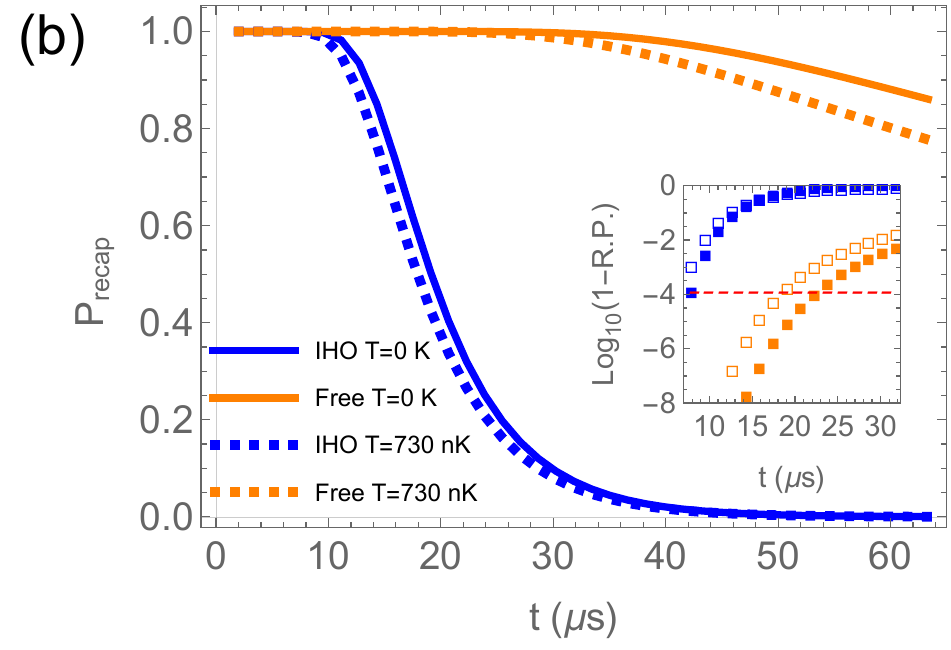}
    \end{minipage}%

\caption{(a) Recapture probabilities of the 1D Free, IHO and IGauss potentials at $T=0$K, for $U_0=50$ $\mu$K and $\omega=25$ kHz. The Free and IHO values are calculated analytically using the propagators of Sec.~\ref{sec:propagators}. The IGauss solutions are calculated numerically. The dashed red lines in the insets indicates the recapture probability bound for $\tau_{\text{recap}}$. The IHO is seen to be a relatively tight lower bound approximation of the IGauss recapture probabilities. (b) Recapture probabilities of the Free and IHO potentials at $T=0$K and $T=730$ nK, for $U_0=50$ $\mu$K and $\omega=25$ kHz. Higher temperatures lower the recapture probabilities, however for realistic temperatures this effect is relatively small.}
    \label{fig:survivalrates}
    \end{centering}
\end{figure}
    \end{minipage}
\end{widetext2}

\begin{widetext2}
    \begin{minipage}[b]{\linewidth}
\begin{figure}[H]
\begin{centering}
\captionsetup{justification=justified}
    \begin{minipage}[t]{.49\textwidth}
         \hypertarget{fig:trecapgaussa}{}
         \includegraphics[width=\columnwidth]{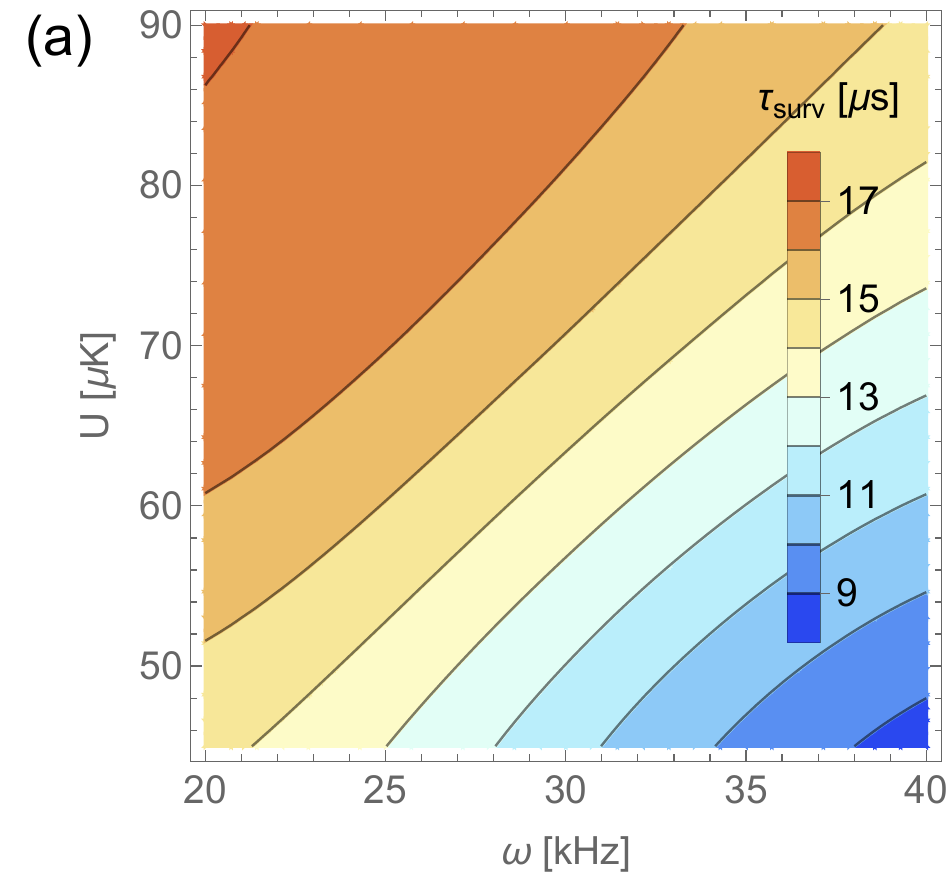}
    \end{minipage}%
    \begin{minipage}[t]{.49\textwidth}
         \hypertarget{fig:trecapgaussb}{}
         \includegraphics[width=\columnwidth]{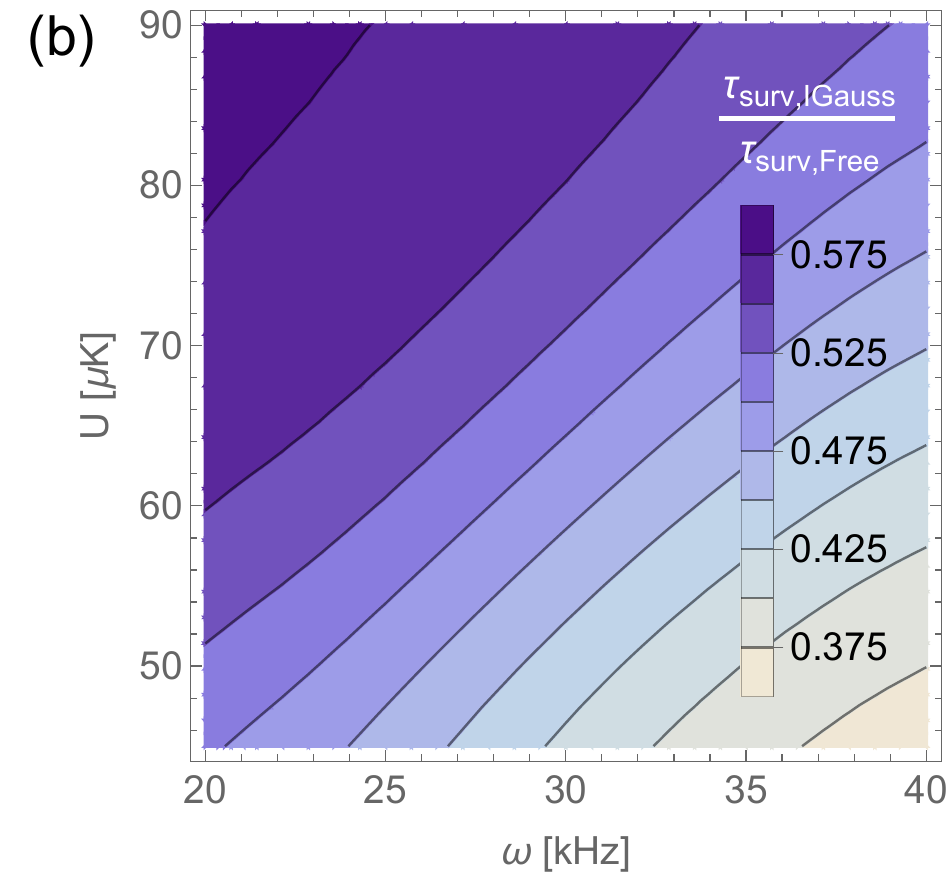}
    \end{minipage}%

\caption{a) Recapture times $\tau_{\text{recap,IGauss}}$ and b) recapture time ratio $\tau_{\text{recap,IGauss}}/\tau_{\text{recap,Free}}$ for varying $U_0$ and $\omega$ in the 1D case. The recapture times increase with increasing $U_0$ and decreasing $\omega_0$. Since contour lines are relatively straight, the ratio $U_0/\omega$ and, through Eq.~\ref{eq:trapedge}, the trap edge $X_{\text{edge}}$ can thus be used as predictors of the recapture times. From b) we note that the recapture times for the IGauss potential stay in the same order of magnitude as the Free potential.}
    \label{fig:trecapgauss}
    \end{centering}
\end{figure}
    \end{minipage}
\end{widetext2}

\begin{figure}[H]
    \centering
    \captionsetup{justification=justified}
    \includegraphics{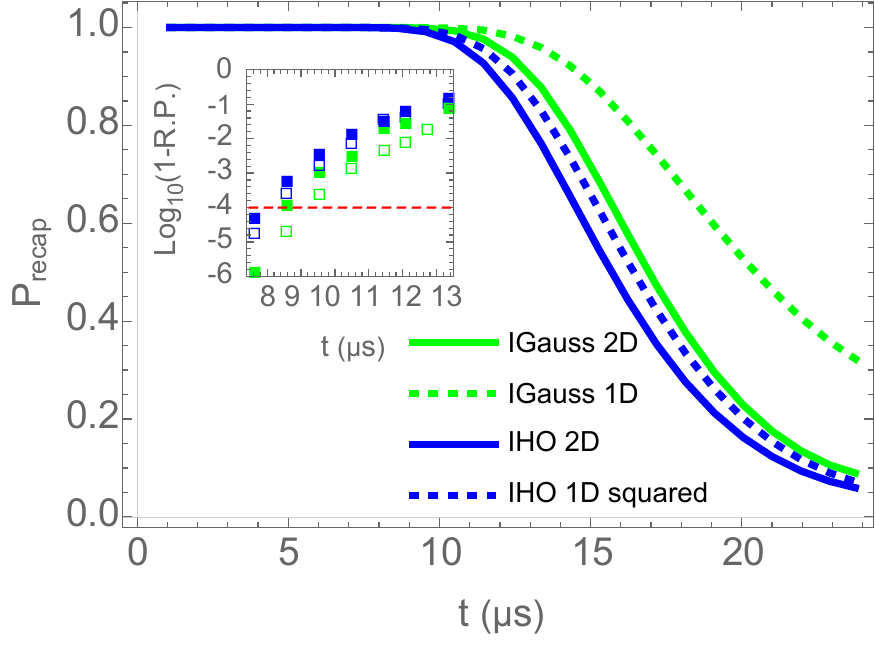}
    \caption{Recapture probabilities for $U_0=50$ $\mu$K and $\omega=25$ kHz for 2D (radial) under IGauss potential, together with the IHO 2D lower bound, and the IHO 1D squared approximation. The dashed red line in the inset indicates the recapture probability bound for $\tau_{\text{recap}}$. The IHO 2D is a lower bound for the IGauss 2D case, and the IHO 1D squared finely approximates the IGauss 2D case.}
    \label{fig:2dresults}
\end{figure}

Figure~\ref{fig:2dresults}, shows the recapture probabilities for the 2D case, compared to the squared 1D approximation. We see indeed that the squared 1D approximation follows the 2D recapture probability. As we are mostly interested in the point $\tau_{\text{recap}}$, this approximation is satisfactory. 

\section{Conclusion}
This work outlines the development of a robust quantum mechanical model for the calculation of recapture probabilities under various potentials. The squared recapture probabilities of a 1D IHO potential (for which an analytical expression exists) nicely approximate the 2D IGauss potential recapture probabilities. From the results of Sec.~\ref{sec:results}, we see that for relevant trap parameters, the recapture time $\tau_{\text{recap}}$ is of the same order of magnitude under a Free potential as under an IGauss potential. This would indicate that for a large array of atoms, it is more beneficial to keep the traps on, thus heating up only a single atom, than to heat up the entire array when switching the traps off.\\

Future studies could incorporate other loss rates, such as black body radiation, stimulated emission in the Rydberg state, and trap scattering caused by off-resonant excitations. Furthermore, the consideration of reachable atom temperatures given the trap parameters $U_0$ and $\omega$ should be taken into account. A treatment of bottle beam traps \cite{bottletraps} using this quantum mechanical model, would also be of great interest.\\

\newpage

\section*{Acknowledgements}
We thank Madhav Mohan, Deon Janse van Rensburg, Jasper Postema, Luke Visser, and Jasper van de Kraats for fruitful discussions. This research is financially supported by the Dutch Ministry of Economic Affairs and Climate Policy (EZK), as part of the Quantum Delta NL program, and by the Netherlands Organisation for Scientific Research (NWO) under Grant No.\ 680.92.18.05.

\section*{Data Availability}
The data that support the findings of this study are available from the corresponding author upon reasonable request.

\bibliographystyle{apsrev4-1}
\bibliography{Bibliography}

\newpage
\onecolumngrid

\appendix
\section{Integrals}
\label{sec:integrals}
For the approximation of the bound states as in Sec.~\ref{sec:section2a}, one needs to construct the matrix with matrix elements $\langle \psi_{\text{HO},n}|H_{\text{Gauss}}|\psi_{\text{HO},m}\rangle$. Numerical integration algorithms show significant errors at high values of $n$ and $m$. Therefore, analytic expressions are used, as
\begin{equation}
\begin{aligned}
    \langle \psi_{\text{HO},n}|H_{\text{Gauss}}|\psi_{\text{HO},m}\rangle=&   \left\langle \psi_{\text{HO},n}\left|H_{\text{HO}}-\frac{1}{2}x^2+V_{\text{Gauss}}\right|\psi_{\text{HO},m}\right\rangle\\
    =&\langle \psi_{\text{HO},n}|H_{\text{HO}}|\psi_{\text{HO},m}\rangle-\left\langle\psi_{\text{HO},n}\left|\frac{1}{2}x^2\right|\psi_{\text{HO},m}\right\rangle+\langle\psi_{\text{HO},n}|V_{\text{Gauss}}|\psi_{\text{HO},m}\rangle.
\end{aligned}
\end{equation}
The first two terms are fairly standard and can be found in most introductory quantum mechanics textbooks \cite{griffiths}. The last term involves terms of the form \cite{hermiteintegrals}
\begin{equation}
\begin{aligned}
    \int_{-\infty}^{\infty} e^{-a x^2} H_n(x) H_m(x) d x=\sqrt{\frac{\pi}{a}} n ! m ! \sum_{k=0}^{\min (n, m)} \frac{2^k(n+m-2 k) !}{k !(n-k) !(m-k) !\left(\frac{n+m}{2}-k\right) !}\left(\frac{1-a}{a}\right)^{\frac{n+m}{2}-k}.
    \end{aligned}
\end{equation}
For the 2D case, the same decomposition is applied to calculate
\begin{equation}
    \langle \psi_{\text{HO},n_1,l_1}|H_{\text{HO}}|\psi_{\text{HO},n_2,l_2}\rangle=(n_1+1)\delta_{n_1=n_2}\delta_{l_1=l_2}.
\end{equation}

For the other two terms
\begin{equation}
\begin{aligned}
    \left\langle\psi_{\text{HO},n}\left|\frac{1}{2}r^2\right|\psi_{\text{HO},m}\right\rangle=&C_0\int_{0}^{2\pi}\int_{0}^\infty r e^{-\frac{r^2}{2}}r^{|l_1|}L_{\alpha_1}^{|l_1|}(r^2)e^{-l_1\phi}\frac{1}{2}r^2e^{-\frac{r^2}{2}}r^{|l_2|}L_{\alpha_2}^{|l_2|}(r^2)e^{+il_2\phi}dr d\phi\\
    =& \frac{1}{2}\pi C_0 \delta_{l_1=l_2}\int_{0}^\infty x^{\frac{|l_1|+|l_2|}{2}+1}e^{-x} L_{\alpha_1}^{|l_1|}(x)L_{\alpha_2}^{|l_2|}(x)dx,\\
    \langle\psi_{\text{HO},n}|V_{\text{Gauss}}|\psi_{\text{HO},m}\rangle=&C_1\int_{0}^{2\pi}\int_{0}^\infty r e^{-\frac{r^2}{2}}r^{|l_1|}L_{\alpha_1}^{(|l_1|)}(r^2)e^{-l_1\phi}\frac{U}{\hbar\omega}e^{\frac{1}{2}r^2\frac{\hbar\omega}{U}}e^{-\frac{r^2}{2}}r^{|l_2|}L_{\alpha_2}^{(|l_2|)}(r^2)e^{+il_2\phi}dr d\phi\\
    =& \pi C_1 \delta_{l_1=l_2}\int_{0}^\infty x^{\frac{|l_1|+|l_2|}{2}}e^{-x(1-\frac{\hbar\omega}{2U})} L_{\alpha_1}^{(|l_1|)}(x)L_{\alpha_2}^{(|l_2|)}(x)dx,
\end{aligned}
\end{equation}
where $L_\alpha^n$ is the generalized Laguerre polynomial, $C_i\in\mathbb{R}_+$ are normalization constants, and the substitution $r^2\rightarrow x$ was made. These integrals can be calculated using the following identity \cite{laguerreintegrals}:
\begin{equation}
    \int_0^{\infty} t^{a-1} e^{-p t} L_m^\lambda(a t) L_n^\beta(b t) d t=\frac{\Gamma(\alpha)(\lambda+1)_m(\beta+1)_n p^{-\alpha}}{m ! n !} \sum_{j=0}^m \frac{(-m)_j(\alpha)_j}{(\lambda+1)_j j !}\left(\frac{a}{p}\right)^j \sum_{k=0}^n \frac{(-n)_k(j+\alpha)_k}{(\beta+1)_k k !}\left(\frac{b}{p}\right)^k.
\label{eq:laguerreintegral}
\end{equation}
Here $\Gamma$ is the gamma function, and $(a)_i$ is the Pochhammer symbol. Note that these symbols are highly discontinuous at negative integers. Therefore, sufficient care should be employed when calculating these expressions.

\section{Initial Quantum Spread}
\label{sec:decayrate}

We show the derivation of the expression of the initial quantum spread as in Eq.~\eqref{eq:decayrates}. Let $l$ indicate a certain potential and consider the function
\begin{equation}
    P_l(t):=|\langle \psi_0|\psi_l(t)\rangle|^2.
\end{equation}
In order to analyze its behavior, we will need the following relations 
\begin{equation}
\begin{aligned}
    &i\partial_t|\psi_l(t)\rangle=-\frac{1}{2}\frac{\partial^2}{\partial x^2}|\psi_l(t)\rangle+V_l|\psi_l(t)\rangle,\\
    &-\frac{1}{2}\frac{\partial^2}{\partial x^2}|\psi_0\rangle+V_l|\psi_0\rangle=E_0|\psi_0\rangle.
\end{aligned}
\end{equation}
The first derivative can be expressed as
\begin{equation}
    \begin{aligned}
    \dot{P}_l(t)&=\overline{\langle \psi_0|\partial_t\psi_l(t)\rangle}\langle\psi_0|\psi_l(t)\rangle+\overline{\langle\psi_0|\psi_l(t)\rangle}\langle \psi_0|\partial_t\psi_l(t)\rangle\\
    &=i\overline{\langle\psi_0|-\frac{1}{2}\frac{\partial^2}{\partial x^2}+V_l|\psi_l(t)\rangle}\langle\psi_0|\psi_l(t)\rangle-i\overline{\langle\psi_0|\psi_l(t)\rangle}\langle\psi_0|-\frac{1}{2}\frac{\partial^2}{\partial x^2}+V_l|\psi_l(t)\rangle\\
    &=i(\langle\psi_l(t)|-\frac{1}{2}\frac{\partial^2}{\partial x^2}+\frac{1}{2}x^2|\psi_0\rangle+\langle\psi_l(t)|V_l-\frac{1}{2}x^2|\psi_0\rangle)\langle\psi_0|\psi_l(t)\rangle\\
    &\,-i\langle\psi_l(t)|\psi_0\rangle(\langle\psi_0|-\frac{1}{2}\frac{\partial^2}{\partial x^2}+\frac{1}{2}x^2|\psi_l(t)\rangle+\langle\psi_0|V_l-\frac{1}{2}x^2|\psi_l(t)\rangle)\\
    &=i\langle\psi_l(t)|E_0|\psi_0\rangle\langle\psi_0|\psi_l(t)\rangle-i\langle\psi_l(t)|\psi_0\rangle\langle\psi_0|E_0|\psi_l(t)\rangle+i\big\langle\psi_l(t)\big|\big[V_l-\frac{1}{2}x^2,|\psi_0\rangle\langle \psi_0|\big]\big|\psi_l(t)\big\rangle\\
    &=i\big\langle\psi_l(t)\big|\big[V_l-\frac{1}{2}x^2,|\psi_0\rangle\langle \psi_0|\big]\big|\psi_l(t)\big\rangle.
    \end{aligned}
\end{equation}
Note that at $t=0$ this object is always equal to zero. For the second derivative, we get
\begin{equation}
    \begin{aligned} 
    \ddot{P}_l(t)& =i\big\langle\partial_t\psi_l(t)\big|\big[V_l-\frac{1}{2}x^2,|\psi_0\rangle\langle \psi_0|\big]\big|\psi_l(t)\big\rangle+i\big\langle\psi_l(t)\big|\big[V_l-\frac{1}{2}x^2,|\psi_0\rangle\langle \psi_0|\big]\big|\partial_t\psi_l(t)\big\rangle\\
    &=i\big\langle\psi_l(t)\big|\left(\frac{1}{2}\frac{\partial^2}{\partial x^2}-V_l\right)\big[V_l-\frac{1}{2}x^2,|\psi_0\rangle\langle \psi_0|\big]\big|\psi_l(t)\big\rangle-i\big\langle\psi_l(t)\big|\big[V_l-\frac{1}{2}x^2,|\psi_0\rangle\langle \psi_0|\big]\left(\frac{1}{2}\frac{\partial^2}{\partial x^2}-V_l\right)\big|\psi_l(t)\big\rangle\\
    &=i\big\langle\psi_l(t)\big|\left[\frac{1}{2}\frac{\partial^2}{\partial x^2}-V_l,\big[V_l-\frac{1}{2}x^2,|\psi_0\rangle\langle \psi_0|\big]\right]\big|\psi_l(t)\big\rangle.
    \end{aligned}
\end{equation}
Looking at $t=0$, we get
\begin{equation}
    \begin{aligned}
    \ddot{P}_l(0)=&i\big\langle\psi_0\big|\left(\frac{1}{2}\frac{\partial^2}{\partial x^2}-V_l\right)\big(\big(V_l-\frac{1}{2}x^2\big)|\psi_0\rangle\langle \psi_0|-|\psi_0\rangle\langle \psi_0|\big(V_l-\frac{1}{2}x^2\big)\big)\\
    &-\big(\big(V_l-\frac{1}{2}x^2\big)|\psi_0\rangle\langle \psi_0|-|\psi_0\rangle\langle \psi_0|\big(V_l-\frac{1}{2}x^2\big)\big)\left(\frac{1}{2}\frac{\partial^2}{\partial x^2}-V_l\right)|\psi_0\rangle\\
    =&\langle \psi_0|\left(\frac{1}{2}\frac{\partial^2}{\partial x^2}-V_l\right)\big(V_l-\frac{1}{2}x^2\big)|\psi_0\rangle-\langle \psi_0|\left(\frac{1}{2}\frac{\partial^2}{\partial x^2}-V_l\right)|\psi_0\rangle\langle\psi_0|\big(V_l-\frac{1}{2}x^2\big)|\psi_0\rangle\\
    &-\langle \psi_0|\big(V_l-\frac{1}{2}x^2\big)|\psi_0\rangle\langle\psi_0|\left(\frac{1}{2}\frac{\partial^2}{\partial x^2}-V_l\right)|\psi_0\rangle+\langle \psi_0|\big(V_l-\frac{1}{2}x^2\big)\left(\frac{1}{2}\frac{\partial^2}{\partial x^2}-V_l\right)|\psi_0\rangle\\
    =&-2\langle\psi_0|V_l^2|\psi_0\rangle+2\langle\psi_0| V_l|\psi_0\rangle^2+2\langle\psi_0|V_l\frac{1}{2}\frac{\partial^2}{\partial x^2}|\psi_0\rangle+2\langle\psi_0| \frac{1}{2}x^2V_l|\psi_0\rangle+2\langle\psi_0|-\frac{1}{4}x^2\frac{\partial^2}{\partial x^2}|\psi_0\rangle\\
    &+2\langle\psi_0|\frac{1}{2}x^2|\psi_0\rangle\langle \psi_0|\frac{1}{2}\frac{\partial^2}{\partial x^2}|\psi_0\rangle-2\langle\psi_0|\frac{1}{2}x^2|\psi_0\rangle\langle \psi_0|V_l|\psi_0\rangle\\
    &-2\langle\psi_0|V_l|\psi_0\rangle\langle \psi_0|\frac{1}{2}\frac{\partial^2}{\partial x^2}|\psi_0\rangle+\langle\psi_0|\frac{1}{2}\frac{\partial^2 V_l}{\partial x^2}|\psi_0\rangle-\langle \psi_0|\frac{1}{2}|\psi_0\rangle-2\langle\psi_0|V_l^2|\psi_0\rangle+2\langle\psi_0| V_l|\psi_0\rangle^2\\
    =&2\left(\frac{1}{2}\langle\psi_0|\big\{\frac{1}{2}\frac{\partial^2}{\partial x^2}+\frac{1}{2}x^2,V_l\}|\psi_0\rangle-\langle \psi_0|\frac{1}{2}\frac{\partial^2}{\partial x^2}+\frac{1}{2}x^2|\psi_0\rangle\langle\psi_0|V_l|\psi_0\rangle\right)\\
    &+2\left(\frac{1}{2}\langle\psi_0|\big\{\frac{1}{2}\frac{\partial^2}{\partial x^2},-\frac{1}{2}x^2\}|\psi_0\rangle-\langle \psi_0|\frac{1}{2}\frac{\partial^2}{\partial x^2}|\psi_0\rangle\langle\psi_0|-\frac{1}{2}x^2|\psi_0\rangle\right).
    \end{aligned}
\end{equation}
We define
\begin{equation}
\begin{aligned}
    \text{Var}_\psi(A)&:=\langle\psi|A^2|\psi\rangle-\langle\psi|A|\psi\rangle^2,\\
    \text{Cov}_\psi(A,B)&:=\frac{1}{2}\langle\psi|\big\{A,B\big\}|\psi\rangle-\langle\psi|A|\psi\rangle\langle\psi|B|\psi\rangle,
\end{aligned}
\end{equation}
and note that $\text{Cov}_\psi(A,A)=\text{Var}_\psi(A)$. We then have
\begin{equation}
     \ddot{P}_l(0)=-2\text{Var}_{\psi_0}(V_l)+2\text{Cov}_{\psi_0}\left(\frac{1}{2}\frac{\partial^2}{\partial x^2}+\frac{1}{2}x^2,V_l\right)-2\text{Cov}_{\psi_0}\left(\frac{1}{2}\frac{\partial^2}{\partial x^2},\frac{1}{2}x^2\right).
\end{equation}
Using linearity of the covariance, we can also write this as 
\begin{equation}
     \ddot{P}_l(0)=-2\text{Cov}_{\psi_0}\left(V_l-\frac{1}{2}\frac{\partial^2}{\partial x^2},V_l-\frac{1}{2}x^2\right).
\end{equation}
We can now calculate this term for the IHO and the IGauss potentials, with $|\psi_0\rangle$ being the HO ground state to get $\ddot{P}_{\text{init,IHO}}(0)=-1$ and $\ddot{P}_{\text{init,IGauss}}(0)=-0.98$ ($U_0=50$ $\mu$K and $\omega=25$ kHz),  indicating faster decay for the IHO than for the IGauss. 

\end{document}